\documentclass[3p,times]{elsarticle}

\usepackage{ecrc}

\usepackage{epstopdf}

\volume{00}

\firstpage{1}

\journalname{Nuclear Physics A}

\runauth{D. Sexty}

\jid{nupha}

\jnltitlelogo{Nuclear Physics A}

\usepackage{graphicx}
\usepackage{amsmath,amssymb}
\newcommand{\bea}{\begin{eqnarray}}
\newcommand{\eea}{\end{eqnarray}}

\begin{document}

\begin{frontmatter}



\title{Progress in complex Langevin simulations of full QCD \\ at nonzero density}

%% Single author (and collaboration) - please insert
\author{D\'enes Sexty}
%\fntext[col1] {A list of members of the XYZ Collaboration and acknowledgements can be found at the end of this issue.}
\address{Institut f\"ur Theoretische Physik, Universit\"at Heidelberg, Heidelberg, Germany}

%% For multiple authors, replace the above by:

%\author[label1]{Author1}
%\author[label2]{Author2}

%\address[label1]{Address 1}
%\address[label2]{Address 2}

\begin{abstract}
Progress in the application of the complex Langevin method to full QCD at non-zero chemical potential is reported. The method evades the sign problem which makes naive simulations at nonzero density impossible. The procedure 'gauge cooling' is used to stabilize the simulations at small enough lattice spacings. The method allows simulations also at high densities, all the way up to saturation. Simulations in a systematic hopping parameter expansion are also performed and good convergence is observed, validating the full as well as the expanded simulations.

%A template for preparing contributions to the proceedings of Quark Matter 2014. The file should be compiled
%with {\em pdflatex}. Figures can be pdf or eps, as illustrated in Fig.~\ref{fig:generic}. If the conversion eps $\to$ pdf
%does not work automatically on your system, you can convert eps files to pdf using a tool like {\em epstopdf}. For special options see\\ 
%\verb!http://www.elsevier.com/author-schemas/preparing-crc-journal-articles-with-latex!.
\end{abstract}


\end{frontmatter}

% \linenumbers


\section{Introduction}
\label{intro}
In recent years, we have seen the great success of lattice QCD, a
 first-principles non-perturbative approach with well 
controlled approximations.
However, the naive application of this method requires that we formulate 
the theory using a path integral with a positive definite measure, 
\bea 
 Z(\mu) = \int DU e^{-S_g[U]} \det M(\mu,U)^{N_F} , 
\eea
where  $S_g[U]$ is the gauge action and $M(\mu,U)$ is the fermion matrix  
which describes $N_F$ quark degrees of freedom. For nonzero 
chemical potential $\mu>0$, the fermionic determinant is 
non positive definite, therefore importance sampling methods are not applicable.
This is known as the 'sign problem'. 
Various methods have been invented to 
negate the sign problem, such as reweighting, Taylor expansion, analytical continuation from imaginary chemical potential, etc. but these are 
mostly successful in the low chemical potential region, 
$ \mu/T \lesssim 1$ (for a review, see \cite{deForcrand:2010ys,Aarts:2013bla}).

 Recent progress in  complex Langevin (CL) dynamics has shown \cite{Seiler:2012wz,Sexty:2013ica} that one 
can evade the QCD sign problem by the complexification of the variables. This 
method allows direct simulations of a theory with complex action 
without any sign or overlap problem (see e.g. applications to Bose 
gas \cite{bosegas}, 
Yang-Mills theory with $ \Theta$-term \cite{thetaterm}, 
real-time physics \cite{realtime}).
The complex Langevin simulations are based on 
on setting up a stochastic process on the complexification of the 
original field manifold \cite{parisi,klauder}. The averages of the 
original theory are recovered using the 
principle of analytic continuation. In some cases, the process is known 
to produce wrong results. The understanding of this behavior has progressed
in the recent years: one can formally prove the correctness of the 
approach \cite{Aarts:2009uq}, provided a few conditions are met, such 
as the holomorphycity of the action and the fast falloff 
of the distribution of the variables in the complexified configuration space.
In the case of QCD, the action we simulate 
$ S_\textrm{eff}[U]=S_{g}[U]+ \textrm{ln det} M(\mu,U) $
is non-holomorphic, it has a branch cut. 
This leads to a meromorphic drift term. Poles in the drift may lead to 
wrong convergence of the process, as shown in nontrivial, soluble
models \cite{Mollgaard:2013qra}. Recent 
evidence suggests that in the case of QCD, at least in some region of 
the parameters, this non-analiticity does not lead to problems,
as seen by comparing two versions of the theory, the full and the 
expanded in which there are no singularities (see below 
and in \cite{Aarts:2014bwa}).

An essential ingredient of the complex Langevin simulations of gauge
theories is gauge cooling \cite{Seiler:2012wz,Aarts:2013uxa}. 
The complexification of the field manifold from SU(3) to SL(3,C) 
gives rise to problems in the naive application of the 
complex Langevin equation, as the link variables try to explore the 
complexified, non-compact manifold SL(3,C). This undesirable 
behavior can be countered by using  
 the gauge freedom in the complexified manifold to move the configuration
closer to the original SU(3) manifold. This is achieved by decreasing the 
'unitarity norm' 
$ \sum\limits_{x,\nu} \textrm{Tr} ( U_{x,\nu} U_{x,\nu}^+ ) $
using gauge transformations, i.e. searching for 
the minimum of the unitarity norm by changing the configuration 
in the direction of the steepest descent.  The dynamical updates are 
interspersed with gauge cooling steps which 
keep the complexified dynamics stable.

One observes that gauge cooling succeeds in controlling the process in 
the complexified field manifold such that a stable behavior is seen, 
without dangerous 'skirts' in the distribution of the variables,  if the 
$\beta$ parameter of the action is not too small. The limiting value 
seems to depend very mildly on the lattice size  \cite{Aarts:2013nja}, so 
in practice 
this limitation means that there is an upper limit on the allowed lattice
spacings that one can use. This value is around $ a_{max}=0.1-0.2$ fm, 
depending on the fermion content of the theory. Lower temperatures
are thus more expensive to simulate, as they require larger lattices.

\begin{figure}
\begin{center}
\includegraphics*[width=10.1cm]{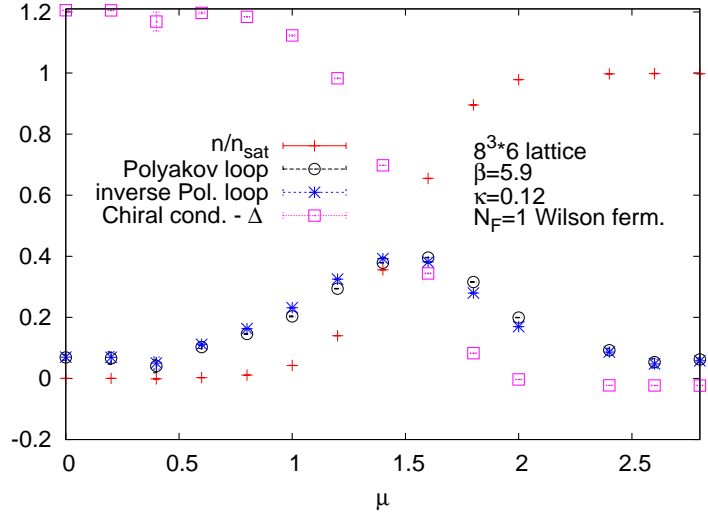}
\caption{ The fermionic density 
 ($ (1/ N^3 N_t) (\partial \textrm{ln} Z / \partial \mu) $, in units of 
the saturation density),
the Polyakov loops, and the chiral condensate $ \langle \bar\psi \psi \rangle$
(shifted by an arbitrary value for better visibility)
in full QCD using 1 flavor of Wilson fermion at fixed temperature as a 
function of the chemical potential.}
\label{wilson_horiz}
\end{center}
\end{figure}

Besides full QCD, we also study an approximation to QCD called 
HDQCD in which the spatial 
hopping parameter $ \kappa_s$ is set to zero.
This is formally justified in the double limit 
$\kappa \rightarrow 0$, $\mu \rightarrow \infty$,\textrm{ with }  $\zeta \equiv 2\kappa e^{\mu} \;\; {\rm fixed}$ \cite{Bender:1992gn}. HDQCD represents 
the leading order (LO) in a systematic expansion of the 
fermionic determinant. This has been extended to next-to-leading order (NLO) 
using the loop expansion \cite{Bender:1992gn,loopexp},
and also to higher orders in combination with the strong coupling 
expansion \cite{strongcoupling}. However, 
going to higher orders is difficult, as one has to consider possible 
fermionic loops and their combinatorial factors. 
 We recently presented an expansion which allows systematic 
 calculation of fermionic corrections to all 
orders \cite{Aarts:2014bwa}, using the full gauge action 
with CL dynamics to negate the remaining sign problem at nonzero 
chemical potential. This approach shows explicitly the 
convergence of the expansion, therefore it validates also the full 
theory where theoretical understanding of the meromorphic drift is still
lacking.

\section{Results}

\begin{figure}
\begin{center}
\includegraphics*[width=8.1cm]{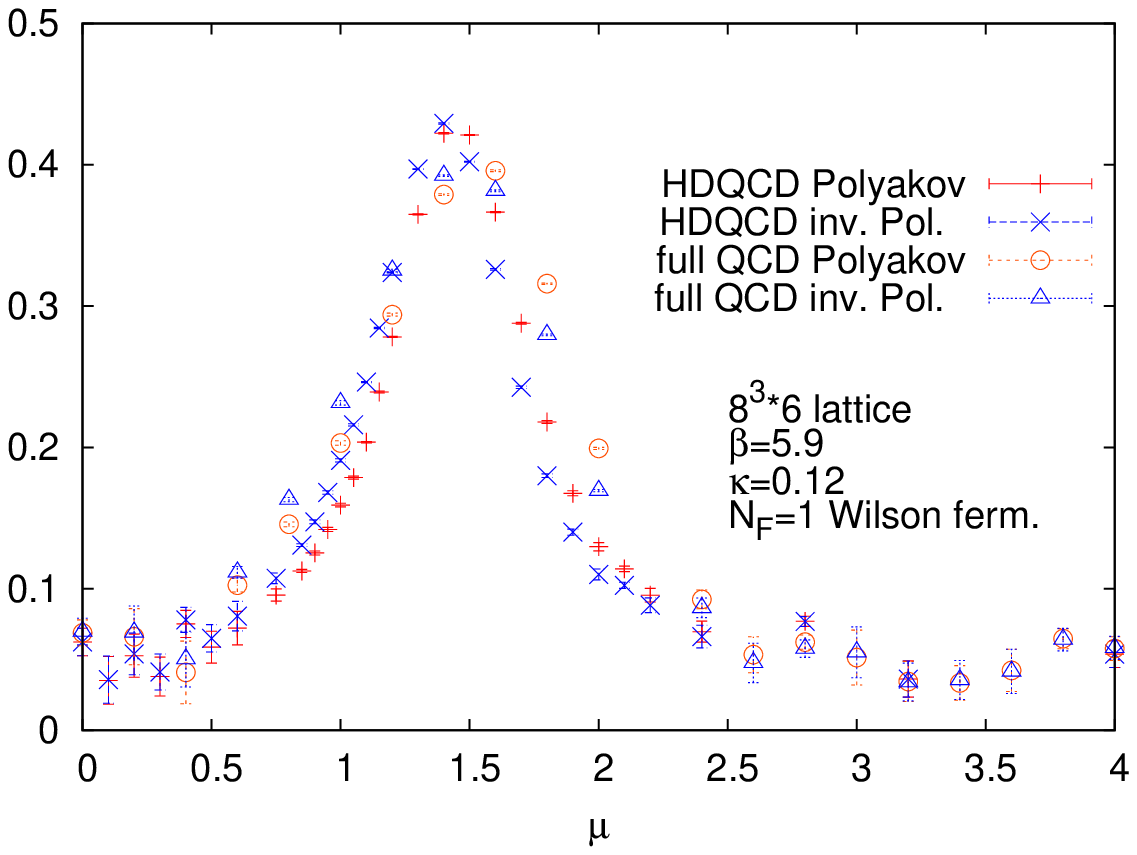}
\includegraphics*[width=8.1cm]{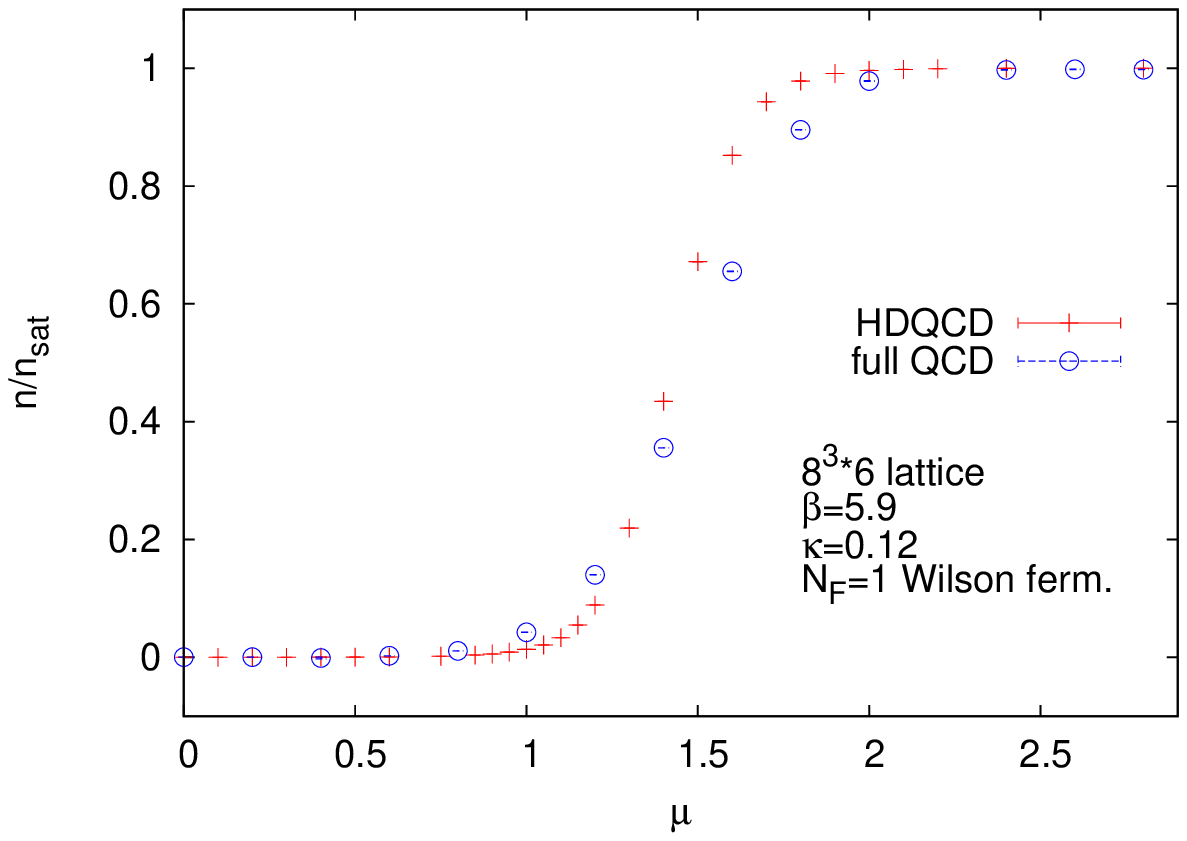}
\caption{
Comparison of the Polyakov loops (left) and fermion density (right) 
in HDQCD and full QCD using 1 flavor of Wilson fermion.}
\label{HDcomp_wilson}
\end{center}
\end{figure}

The method has been implemented for staggered fermions in \cite{Sexty:2013ica}. 
Here results using unimproved Wilson fermions are presented, 
using the fermion matrix
\bea
M(x,y)= 1 - \sum\limits_{\nu=1}^4 \kappa_\nu
\left( (1-\gamma_\nu) \exp ( {\delta_{\nu 4} \mu})
U_\nu(x) \delta_{y,x+a_\nu}  
+ (1 +\gamma_\nu) \exp ( {-\delta_{\nu 4} \mu})
U_\nu^{-1} (y) \delta_{y,x-a_\nu} 
\right),
\eea
with the hopping parameters $\kappa_\nu$ (the spatial hopping
$ \kappa_s =\kappa_1 = \kappa_2 =\kappa_3 $)
 and Euclidean 
Gamma-matrices $ \gamma_\nu$. The contribution of the fermions to the drift
term is calculated using a noisy estimator similarly to the staggered case, 
where the numerical cost is one conjugate gradient solution per update.

The response of the lattice system to the chemical potential 
is demonstrated on Fig.~\ref{wilson_horiz}. The
fermionic density rises to a large value, until all the available 
states on the lattice are filled, and the system is saturated. 
The Polyakov loops signal deconfinement, but decay again when 
lattice effects of the saturation become important, their peaks 
being close to the point of half filling, when every other 
fermionic state is filled on the average.

On Fig.~\ref{HDcomp_wilson} we compare the HDQCD approximation to the 
full QCD using one flavor of Wilson fermion. One observes that 
the qualitative behavior is very similar, both theories exhibit the 
phenomenon of saturation.

\begin{figure}
\begin{center}
\includegraphics*[width=7.5cm]{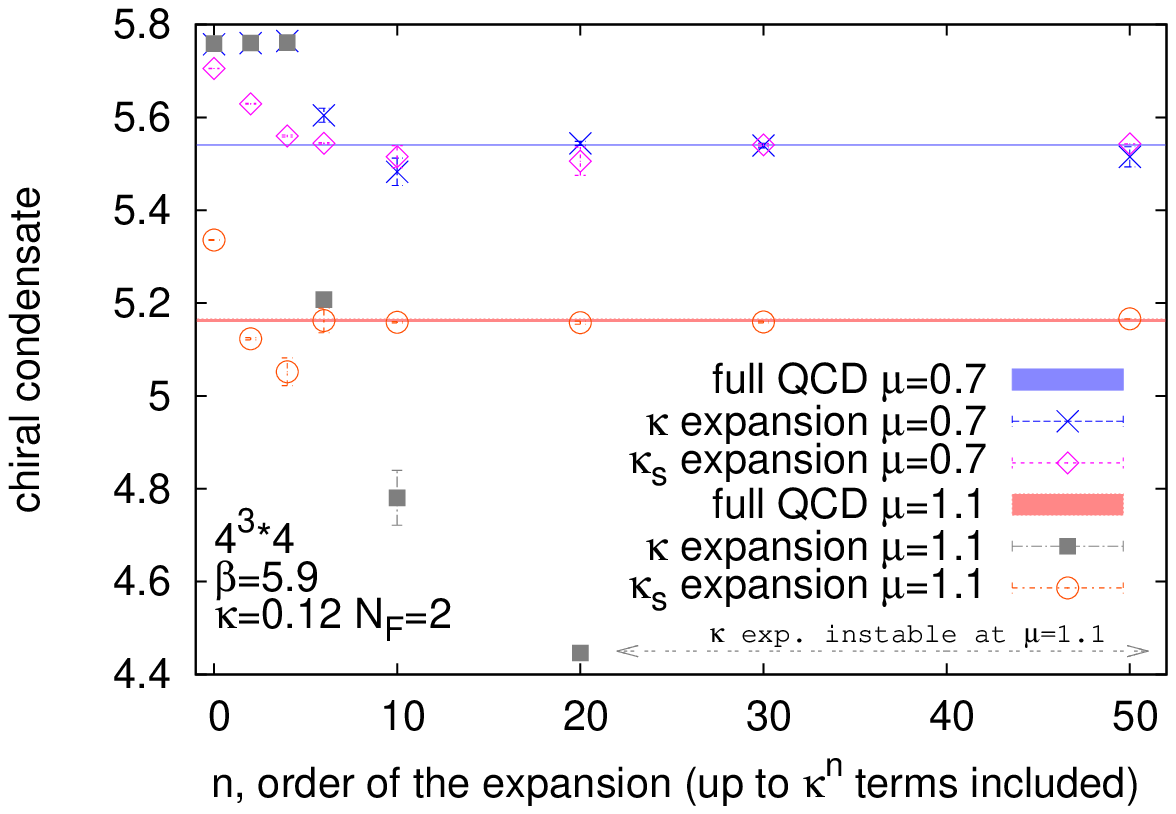}
\includegraphics*[width=7.5cm]{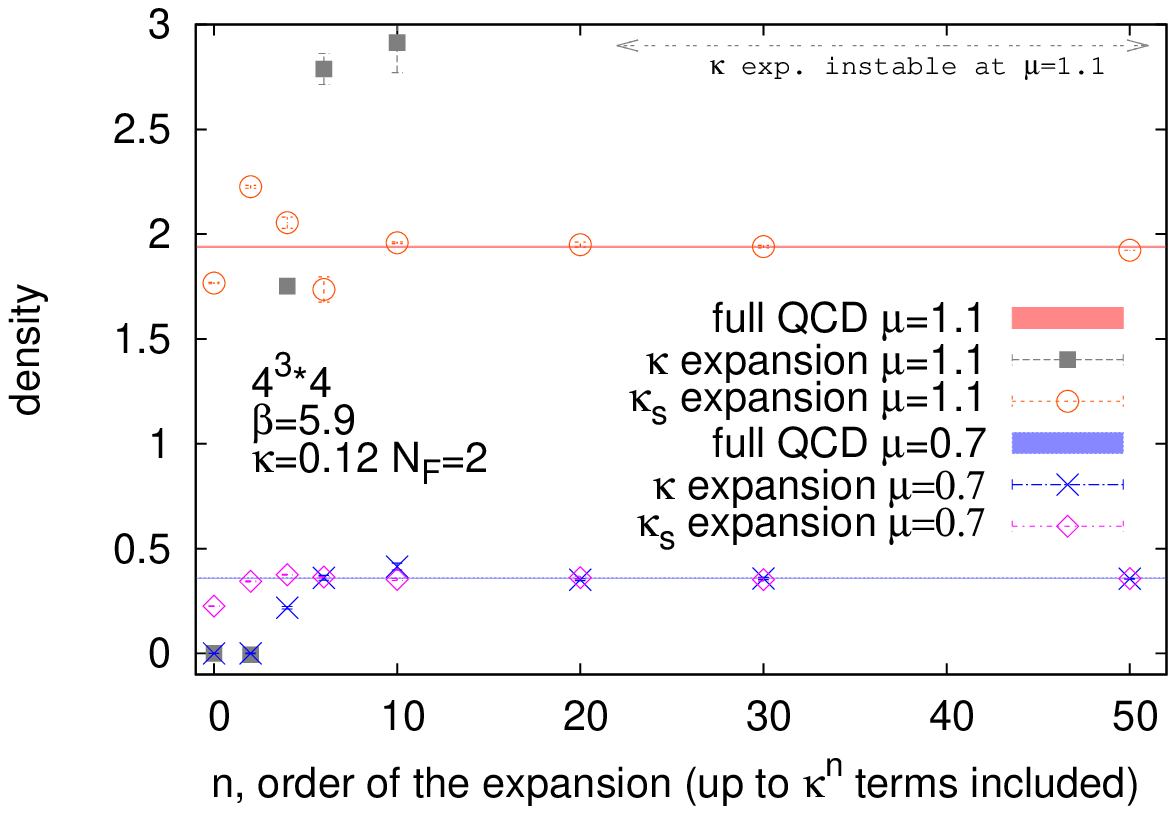}
\caption{ The convergence of the $\kappa$- and $\kappa_s$-expansions
as a function of the order of the corrections taken into account. }
\label{kappasexp}
\end{center}
\end{figure}

We define the $\kappa$-expansion by rewriting the fermionic determinant 
\bea
 \det M = \det(1-\kappa Q) = \exp\sum\limits_{n=1}^{\infty}  
- {\kappa^n \over n} \textrm{Tr}\,Q^n,
\eea 
 where $Q$ is the hopping part of the matrix, and the 
$ \kappa_s$-expansion by first pulling 
out the temporal hopping terms $R$ 
\bea
\det M =\det (1-R - \kappa_s S ) =  \det (1-R) \left( 1 - \frac{1 }{1-R} \kappa_s S  \right)=
\det (1-R) \exp \sum\limits_{n=1}^{\infty}  
- \frac{\kappa_s^n}{n} \textrm{Tr} \left(  \frac{1}{1-R}  S \right)^n,
 \label{kappasdet}
\eea
leaving an expansion in terms of the spatial hopping matrix $S$. 
These expansions can be conveniently implemented in the 
complex Langevin dynamics using noisy estimators \cite{Aarts:2014bwa}.
On Fig.~\ref{kappasexp} the performance of 
these expansions is demonstrated. The $\kappa$-expansion is slightly 
cheaper to calculate, but converges only at small chemical potentials,
as the expansion includes terms which are proportional to $ e^\mu$. 
In contrast, in the $\kappa_s$-expansion the $\mu$ dependent terms 
are pulled out to be dealt with analytically, so it has better convergence properties at large $\mu$, but it is more expensive, as the (analytic) calculation 
of the inverse of the matrix $(1-R)^{-1}$ is needed.

\section{Conclusions}

The complex Langevin method presents a way to evade the sign problem
in theories with complex actions. Promising recent developments
have shown that it delivers very sensible results for QCD-like theories
and even for full QCD. 

At small chemical potentials, where reweighting 
is feasible, the CL approach is validated by checking for 
agreement \cite{Seiler:2012wz,fullcomp}.
The theoretical foundation of the method in the case of a meromorphic action
is not properly understood yet, but agreement with the
systematic $\kappa$- and $\kappa_s$-expansions can be used to validate 
the method, by demonstrating 
that the poles cause no harm to the results. 

These recent results show promise that the Complex Langevin approach will allow 
the exploration of the phase diagram of full QCD in detail, just 
as the phase diagram of HDQCD \cite{ben}.

{\it Acknowledgments} -- I thank to G.~Aarts, F.~Attanasio, Z.~Fodor,
B.~J\"ager, S.D.~Katz, E.~Seiler and I.-O.~Stamatescu 
for discussions and collaboration on related work. 
Parts of the calculations were done on the 
bwGRiD (http://www.bw-grid.de), member of the German D-Grid initiative, 
funded by BMBF and MWFK Baden-W\"urttemberg.



%% References


%\bibliographystyle{elsarticle-num}



\end{document}